\documentclass[aps,prd,amsmath,showpacs,preprintnumbers,superscriptaddress,
               twocolumn,floatfix]{revtex4}
\usepackage{amsmath}
\usepackage{graphicx}
\usepackage{latexsym}
\usepackage{amssymb}

\newcommand{\beqa}{\begin{eqnarray}}
\newcommand{\eeqa}{\end{eqnarray}}
\newcommand{\beqn}{\begin{equation}}
\newcommand{\eeqn}{\end{equation}}

\begin{document}

\title{Infrared behavior of gluon and ghost propagators 
from asymmetric lattices}
\author{Attilio Cucchieri}
\author{Tereza Mendes}
\affiliation{Instituto de F\'\i sica de S\~ao Carlos,
Universidade de S\~ao Paulo, \\
C.P.\ 369, 13560-970, S\~ao Carlos, SP, Brazil}

\begin{abstract} 
\noindent
We present a numerical study of the lattice Landau gluon and ghost propagators
in three-dimensional pure $SU(2)$ gauge theory. Data have been
obtained using asymmetric lattices ($V = 20^2 \times 60$,
$40^2 \times 60$, $8^2 \times 64$,
$8^2 \times 140$, $12^2 \times 140$ and
$16^2 \times 140$) for the lattice coupling $\beta = 3.4$,
in the scaling region. We find that the gluon (respectively ghost) propagator
is suppressed (respec.\ enhanced) at small momenta in the limit of
large lattice volume $V$. By comparing these
results with data obtained using symmetric lattices
($V = 60^3$ and $140^3$), we find that both
propagators suffer from systematic effects in the
infrared region ($p \lesssim 650 \mbox{MeV}$). In particular,
the gluon (respec.\ ghost) propagator is less IR-suppressed 
(respec.\ enhanced) than in the symmetric case.
We discuss possible implications of the use of asymmetric lattices.
\end{abstract}


\maketitle

\section{Introduction}

Gluon and ghost propagators are powerful tools in the
(non-perturbative) investigation of the infra-red (IR) limit of QCD
and are thus of central importance for understanding confinement 
(see e.\ g.\ \cite{Alkofer2000,Holl:2006ni}).
In fact, according to the Gribov-Zwanziger
\cite{Gribov:1977wm,Zwanziger:1993dh,
Sobreiro:2004us,Dudal:2005na,Sobreiro:2005ec}
and to the Kugo-Ojima \cite{Kugo:1979gm,Kugo:1995km}
confinement scenarios in Landau gauge,
the ghost propagator must show a divergent behavior --- stronger than
$p^{-2}$ --- for vanishing momentum $p$.
This strong infrared divergence corresponds to a long-range interaction in
real space, which may be related to quark confinement.
At the same time, according to the former scenario, the gluon propagator
must be suppressed and may go to zero in the IR limit
\cite{Gribov:1977wm,Zwanziger:1993dh,
Sobreiro:2004us,Dudal:2005na,Sobreiro:2005ec,Zwanziger:1990by,Zwanziger:gz,Zwanziger:1991ac}.
This would imply that the real-space gluon propagator
is positive and negative in equal measure, i.e.\ reflection positivity is maximally
violated \cite{Cucchieri:2004mf,Furui:2004cx}, indicating absence of gluons
from the physical spectrum (gluon confinement).
These predictions are also valid for the case of pure $SU(2)$ gauge
theory and for three space-time dimensions.

The nonperturbative evaluation of gluon and ghost propagators can be
achieved by first principles methods such as Dyson-Schwinger equations 
(DSE's) \cite{Alkofer2000,Holl:2006ni}
or the numerical simulation of lattice QCD.
Landau-gauge studies of DSE's (see e.\ g.\
\cite{Zwanziger:2001kw,Lerche:2002ep,Alkofer:2003jj}) have
found an IR behavior of the form $D(p^2) \sim
p^{4 \kappa - 2} = p^{2 a_D - 2} $ [implying $D(0) = 0$ if $\kappa > 0.5$]
for the gluon propagator and of the form
$G(p^2) \sim p^{-2 \kappa -2} = p^{-2 a_G -2}$ for the ghost propagator,
with the same exponent $\kappa$ (i.e.\ with $a_D = 2 a_G$).
The available predictions for the IR exponent point towards
$\kappa \gtrsim 0.5$ for pure $SU(N_c)$ gauge theory in the four dimensional case.
For the $3d$ case the exponents are $a_G \approx 0.4$ and $a_D \approx 1.3$.
Note that in the $d$ dimensional case
\cite{Zwanziger:2001kw,Lerche:2002ep} the relation between $a_D$ and $a_G$
is given by $a_D = 2 a_G + (4 - d)/2$, implying for the quantity 
$\alpha(p^2) = (g^2 / 4 \pi) D(p^2) G^2(p^2) p^6$ the infrared behavior
$p^{2 (a_D - 2 a_G)} = p^{4 - d}$. Thus, in the 4-dimensional case the
running coupling $\alpha(p^2)$ displays and infrared fixed point.

Numerical studies of lattice gauge theories confirm the IR divergence
of the Landau ghost propagator
\cite{Suman:1995zg,Cucchieri:1997dx,Furui:2003jr,Bloch:2003sk} 
and have now also established that the Landau gluon propagator
shows a turnover in the IR region, attaining a finite value 
for $p \approx 0$. (A reliable extrapolation of $D(0)$ to the 
infinite-volume limit is still lacking \cite{Cucchieri:2003di}.)
More precisely, indications of a decreasing gluon propagator 
for small $p$ have been obtained
in the $4d$ $SU(2)$ and $SU(3)$ cases
for the strong-coupling regime
\cite{Cucchieri:1997dx,Cucchieri:1997fy,Nakajima:1998ch},
in the $3d$ $SU(2)$ case (also in the scaling region)
\cite{Cucchieri:1999sz,Cucchieri:2000cy,Cucchieri:2001tw},
in the $3d$ $SU(2)$ adjoint Higgs model
\cite{Cucchieri:2001tw} and
in the $4d$ $SU(2)$ case at finite temperature
\cite{Bogolubsky:2002ui}.
The actual turning of the gluon propagator has been seen clearly
for the equal-time three-dimensional transverse
gluon propagator in $4d$ $SU(2)$ Coulomb gauge
\cite{Cucchieri:2000gu,Cucchieri:2000kw},
for the $3d$ $SU(2)$ Landau case using very large lattices
\cite{Cucchieri:2003di} (of 140 lattice sites per direction)
and --- recently --- in the $4d$ $SU(3)$ Landau case with the use of
asymmetric lattices \cite{Oliveira:2004gy,Silva:2005hd,Oliveira:2005hg,
Silva:2005hb}.

Thus, the two nonperturbative approaches above seem to support the
Gribov-Zwanziger and the Kugo-Ojima confinement scenarios in
Landau gauge. However, the agreement between the two methods is
still at the qualitative level, at least when considering the
gluon and the ghost propagators.
Moreover, recent lattice studies \cite{Boucaud:2005ce,Ilgenfritz:2006gp}
seem to indicate a null IR limit for $\alpha(p^2)$,
instead of a finite nonzero value.
Of course, in this case, special care must be taken to eliminate
finite-size effects, especially when the IR region is considered
\cite{Bloch:2003sk,Bloch:2002we,Cucchieri:1997fy,Cucchieri:1999sz,
Bonnet:2001uh,Cucchieri:2003di}.
This issue has also been studied analitically using
DSE's \cite{Fischer:2005ui}, showing that torus and continuum
solutions are qualitatively different. This suggests a
nontrivial relation between studies on compact and on noncompact manifolds.
Finally, the nonrenormalizability of the ghost-gluon vertex
--- proven at the perturbative level \cite{Taylor:1971ff},
confirmed on the lattice \cite{Cucchieri:2004sq,Ilgenfritz:2006gp}
(for $p \gtrsim 200$ MeV) and used in
DSE studies to simplify the coupled set of equations
--- has been 
recently criticized in Ref.\ \cite{Boucaud:2005ce}. Thus, 
clear quantitative understanding of the two confinement scenarios is
still an open problem.

In this work we extend the study presented in \cite{Cucchieri:2003di}
for the $3d$ $SU(2)$ case,
by including results for the ghost propagators from very large
lattices and by considering also asymmetric lattices,
as done in \cite{Oliveira:2004gy,Silva:2005hd,Oliveira:2005hg,
Silva:2005hb,Leinweber:1998uu,Parappilly:2006si}.
Our aim is to verify possible systematic effects related to the use
of asymmetric lattices (as suggested in \cite{Ilgenfritz:2006gp}), 
by comparing the results to the ones obtained for symmetric 
lattices. In particular, we focus on the determination of the
IR critical exponents $ a_D $ and $ a_G $ and on their extrapolation to
the infinite-volume limit.
The study of gluon and ghost
propagators in three space-time dimensions is computationally
much simpler than in the four-dimensional case and it may help to
get a better understanding of the $4d$
case \cite{Feynman:1981ss,Cornwall:1988ad}.
Note that the $3d$ case is also of interest in
finite-temperature QCD (see for example \cite{Karsch:1998tx,
Maas:2005ym}).

In the following section we describe our simulations and present our
results.
We end with our conclusions.


\section{Simulations and Results}
\label{sec:simula}

Simulations have been done using the standard Wilson action for
$SU(2)$ lattice gauge theory in three dimensions with periodic
boundary conditions. We consider $\beta = 3.4$ and several lattice
volumes, up to $V = 140^3$. In Table \ref{tab:runs} we report, 
for each lattice volume $V$, the parameters used for the simulations.
All our runs start with a random gauge configuration.
We used the {\em hybrid overrelaxed} algorithm (HOR)
\cite{Brown:1987rr,Wolff:1992ri}
for thermalization and the {\em stochastic overrelaxation}
algorithm or the so-called {\em Cornell method} for
gauge fixing thermalized configurations
to the lattice Landau gauge \cite{Cucchieri:1995pn,Cucchieri:1996jm,
Cucchieri:2003fb}.
The numerical code is parallelized using MPI \cite{Cucchieri:2003zx}.
For the random number generator we use a double-precision
implementation of RANLUX (version 2.1) with luxury level set to 2.
Computations were performed on the PC clusters at the IFSC-USP.
We ran part of the $60^3$ lattices on three nodes and the $140^3$
lattices on four or eight nodes.

\begin{table}[b]
\protect\vskip 1mm
\caption{Lattice volumes $V$,
configurations,  HOR sweeps used for
thermalization and between two
configurations (used for evaluating both propagators).}
\label{tab:runs}
\begin{tabular}{ c c c c c c }
$V$ & $\,\;$ Configurations $\,\;$ & $\,\;$ Thermalization $\,\;$ & 
$\,\;$ Sweeps \\ \hline
  $20^3$ & 500 & 550 & 50 \\
  $20^2 \times 40$ & 800 & 440 & 40 \\
  $20^2 \times 60$ & 600 & 550 & 50 \\
  $60^3$ & 400 & 1100 & 100 \\
  $8^2 \times 64$ & 100 & 550 & 50 \\
  $8^2 \times 140$ & 1000 & 550 & 50 \\
  $12^2 \times 140$ & 600 & 660 & 60 \\
  $16^2 \times 140$ & 600 & 770 & 70 \\
  $140^3$ & 101 & 1650 & 150 \\
\end{tabular}
\end{table}

In order to set the physical scale we consider $3d$ $SU(2)$
lattice results for the string tension and the input
value $\sqrt{\sigma} = 0.44 \, GeV$, which is a typical
value for this quantity in the $4d$ $SU(3)$ case.
With $\hbar = c = 1$, this implies
$1 \,fm^{-1} = 0.4485 \sqrt{\sigma}$.
For $\beta = 3.4$ we obtain the average plaquette value
$ \langle W_{1,1} \rangle = 0.672730(3) $. Then, considering
the tadpole-improved coupling $\beta_{I} \equiv \beta\,
\langle W_{1,1} \rangle $, we can evaluate the
string tension $ \sqrt{\sigma} $ in lattice units
using the fit reported in Eq.\ 2 and Table IV of Ref.\
\cite{Lucini:2002wg}. (The fit is valid for $\beta \gtrsim 3.0$,
i.e.\ the coupling $\beta$ considered here is well
above the strong-coupling region.) This gives $ \sqrt{\sigma} =
0.506(5) $, implying a lattice spacing $ a = 0.227(2) $ fm,
i.e.\ $ a^{-1} = 0.869(8) $ GeV. Also,
we are able to consider here momenta as small as
$ p_{min} =  39.0(4) $ MeV
and physical lattice sides of almost 32 fm.
Finally, let us notice that, if we compare the data for the string
tension (in lattice units) with the data reported in Ref.\ \cite{Fingberg:1992ju}
for the $SU(2)$ group in four dimensions (see also Ref.\ \cite{Bloch:2003sk}),
then our value of $\beta$ corresponds to $\beta \approx 2.21$
in the $4d$ case.


We study the lattice gluon and ghost propagators [respectively,
$D(p^2)$ and $G(p^2)$] as a function of the magnitude of
the lattice momentum $ p(k)$
(see Ref.\ \cite{Cucchieri:1999sz,Cucchieri:2005yr} for definitions).
We consider all vectors $k \equiv (k_{x}\,,
k_{y}\,, k_{t})$ with only one component different from zero. For asymmetric
lattices, the non-zero component has been taken along the elongated direction;
for symmetric lattices, when evaluating the gluon propagator, we average
over the three directions in order to increase statistics.
Here, we do not check for possible effects of
the breaking of rotational invariance
\cite{Becirevic:1999uc,Ma:1999kn}. Nevertheless, as we will see
below, systematic effects due to the use of asymmetric
lattices are evident mostly in the IR limit, where one expects
the breaking of the rotational symmetry to play a small effect
(see also Ref.\ \cite{Cucchieri:2003di}).

\begin{figure}[t]
\begin{center}
\protect\vskip -4.0cm
\includegraphics[height=1.35\hsize]{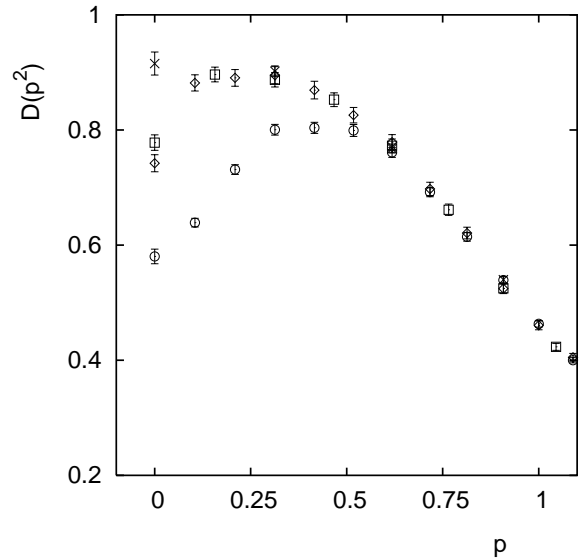}
\protect\vskip -0.3cm
\end{center}
\caption{Plot of the gluon propagator $D(p^2)$ as a function of $p$
(both in lattice units)
for lattice volumes $V = 20^3\,(\times)$, $20^2 \times 40
\, (\Box)$, $20^2 \times 60 \, (\Diamond)$ and $V = 60^3 \, (\bigcirc)$.
}
\label{fig:gluon1}
\end{figure}

\begin{figure}
\begin{center} 
\protect\vskip -4.0cm
\includegraphics[height=1.35\hsize]{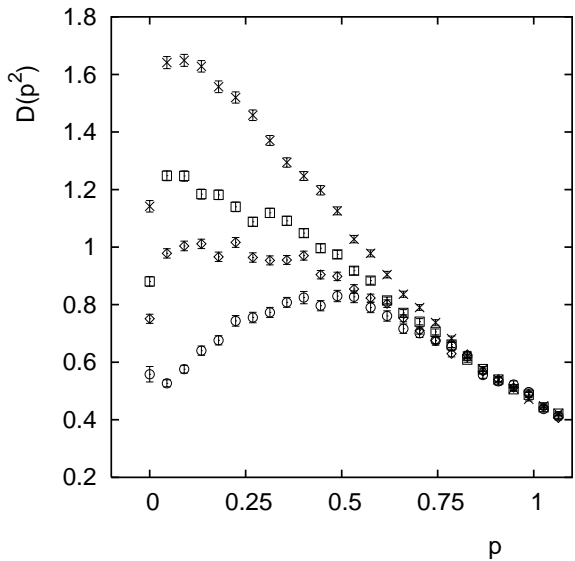}
\protect\vskip -0.3cm
\end{center}
\caption{Plot of the gluon propagator $D(p^2)$ as a function of $p$
(both in lattice units)
for lattice volumes $V = 8^2 \times 140\,(\times)$,
$12^2 \times 140
\, (\Box)$, $16^2 \times 140 \, (\Diamond)$ and $V = 140^3 \,
(\bigcirc)$.
}
\label{fig:gluon2}
\end{figure}

\begin{figure}
\begin{center}
\protect\vskip -4.0cm
\includegraphics[height=1.35\hsize]{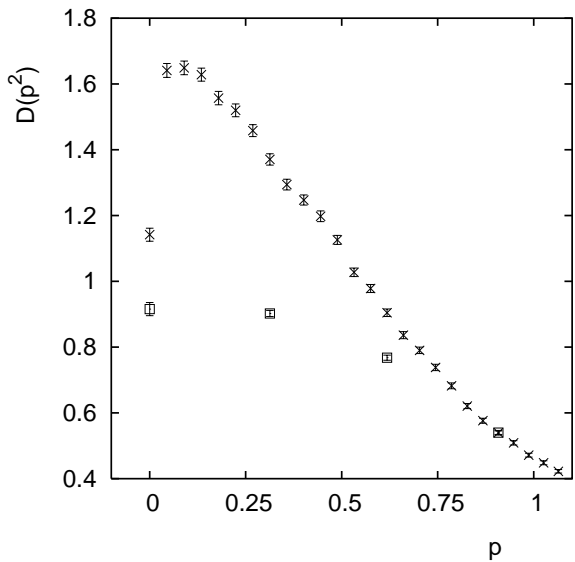}
\protect\vskip -0.3cm
\end{center}
\caption{Plot of the gluon propagator $D(p^2)$ as a function of $p$
(both in lattice units)
for lattice volumes $V = 8^2 \times 140\,(\times)$
and $20^3 \, (\Box)$.
}
\label{fig:gluon3}
\end{figure}

\begin{table}
\protect\vskip 1mm
\caption{Fit of the gluon dressing function $ Z(p^2) = D(p^2) p^2 $ using the fitting function
$z \,(p^2)^{a_D} (1 + a_1 p^2 + a_2 p^4) $ and data
in the interval $p^2 \leq 0.85/a^2$. For each
lattice volume $V$ we report the infrared
exponent $a_D$, the number of degrees of freedom ($d.o.f.$)
and the $\chi^2/d.o.f.$ 
For the lattice volume $V = 20^2$ there were not enough
data points to do the fit.}
\label{tab:fit_gluon}
\begin{tabular}{c c c c}
$V$ & $\;\;\;\;a_D\;\;\;\;$ & $\chi^2 / d.o.f.$ & $ \# d.o.f.$ \\ \hline
$20^2 \times 40\;$ & 1.026 (7) & 0.4 & 2 \\ 
$20^2 \times 60\;$ & 1.028 (5) & 0.4 & 5 \\ 
$60^3\;$ & 1.133 (9) & 1.1 & 5 \\ 
$8^2 \times 64\;$ & 0.94 (4) & 1.2 & 5 \\ 
$8^2 \times 140\;$ & 0.981 (5) & 1.3 & 17 \\ 
$12^2 \times 140\;$ & 0.991 (5) & 1.2 & 17 \\ 
$16^2 \times 140\;$ & 1.019 (6) & 1.2 & 17 \\ 
$140^3\;$ & 1.13 (1) & 1.2 & 17 \\ \hline
\end{tabular}
\protect\vskip 1mm
\end{table}

In Figures \ref{fig:gluon1}, \ref{fig:gluon2} and \ref{fig:gluon3} we
plot the data obtained for the gluon propagator for different lattice
volumes. (Errors represent one standard deviation.)
One can clearly see that the propagator is IR-suppressed
for sufficiently large lattice volumes. This behavior is evident both
for the asymmetric and the symmetric lattices. However, the shape of
the propagator is clearly different in the two cases for momenta $p
\lesssim 0.75/a$, i.e.\ $p \lesssim 0.65 $ GeV. Indeed, for asymmetric lattices
the gluon propagator starts to decrease only at very small momenta
(see Figures \ref{fig:gluon1} and \ref{fig:gluon2}). A similar behavior
can also be observed 
in Fig.\ 2 of Ref.\ \cite{Silva:2005hd}
and in Fig.\ 4 of Ref.\ \cite{Silva:2005hb}.
Thus, considering the asymmetric lattices one would estimate 
a value $M \lesssim 0.25/a \approx 0.22 $ GeV
as a turnover point in the IR, i.e.\ the momentum $p = M$
for which the propagator reaches its peak.
On the other hand, considering the two
largest lattice volumes,
i.e.\ $ V = 60^3 $ (see Fig.\ \ref{fig:gluon1}) and $ V = 140^3 $
(see Fig.\ \ref{fig:gluon2}), the gluon propagator is clearly a decreasing
function of $p$ for $p \lesssim 0.5/a$, corresponding to
$M \lesssim 0.435$ GeV. This is in agreement with Refs.\
\cite{Cucchieri:2003di,Cucchieri:1999sz} where
the value $M = 350^{+100}_{-50} \,MeV$ is reported.
We thus see a difference of almost a factor 2 between the
momentum-turnover point in the symmetric and asymmetric cases.

\begin{figure}
\begin{center}
\protect\vskip -4.0cm
\includegraphics[height=1.35\hsize]{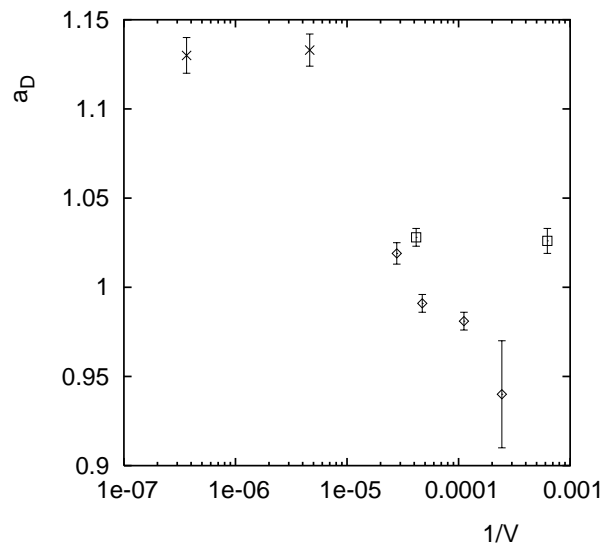}
\protect\vskip -0.3cm
\end{center}
\caption{Plot of the gluon IR exponent $ a_D $ as a function of the inverse
lattice volume $1 / V $ (in lattice units)
for lattice volumes: $V = 20^2 \times 40$ and $20^2 \times 60$ $\,(\Box)$, 
$V = 8^2 \times 64$, $8^2 \times 140$, $12^2 \times 140$ and
$16^2 \times 140$ $\,(\Diamond)$, $V = 60^3$ and $140^3$ $\,(\times)$.
Note the logarithmic scale on the $x$ axis.
}
\label{fig:extrap}
\end{figure}

\begin{figure}
\begin{center} 
\protect\vskip -4.0cm
\includegraphics[height=1.35\hsize]{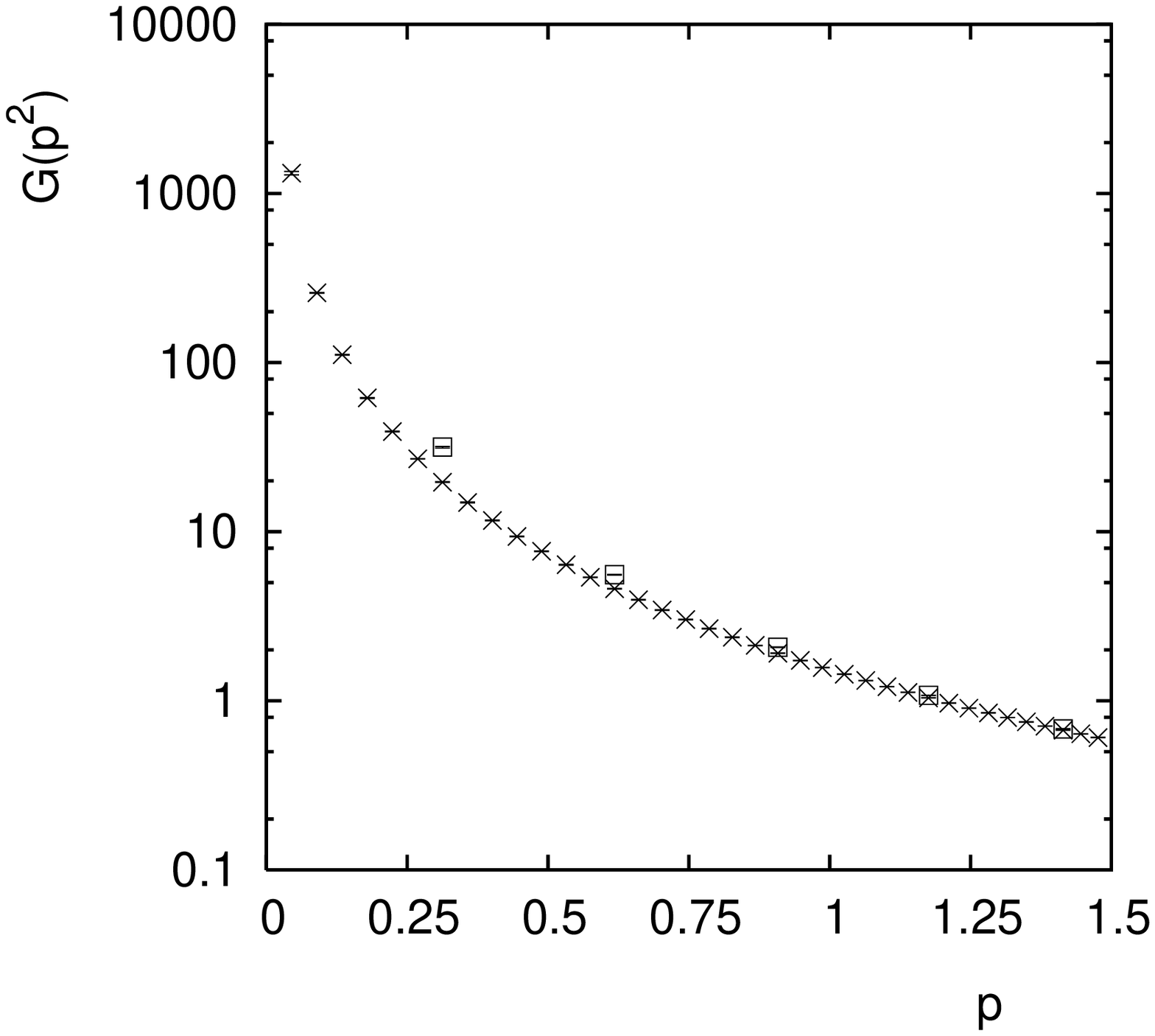}
\protect\vskip -0.3cm
\end{center}
\caption{Plot of the ghost propagator $G(p^2)$ as a function of $p$ 
(both in lattice units)
for lattice volumes $V = 8^2 \times 140\,(\times)$ and $20^3
\, (\Box)$.
Note the logarithmic scale on the $y$ axis.
}
\label{fig:ghost1}
\end{figure}

\begin{figure}[t]
\begin{center}
\protect\vskip -4.0cm
\includegraphics[height=1.35\hsize]{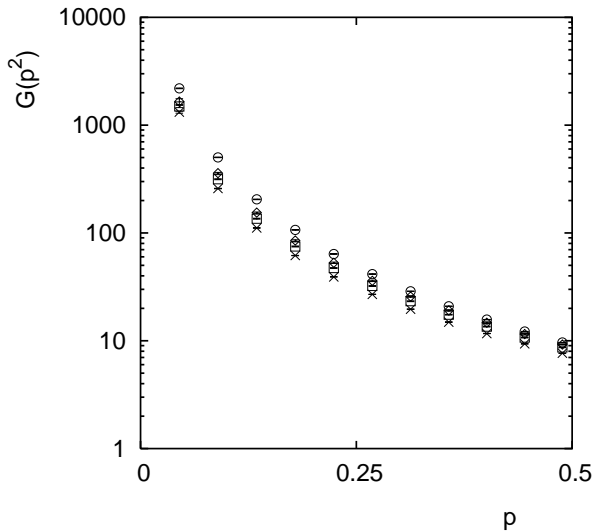}
\protect\vskip -0.3cm
\end{center}
\caption{Plot of the ghost propagator $G(p^2)$ as a function of $p$ 
(both in lattice units)
for lattice volumes $V = 8^2 \times 140\,(\times)$, $12^2 \times 140
\, (\Box)$, $16^2 \times 140 \, (\Diamond)$ and $V = 140^3 \, (\bigcirc)$.
Note the logarithmic scale on the $y$ axis.
}
\label{fig:ghost3}
\end{figure}

\begin{table}
\protect\vskip 1mm
\caption{Fit of the ghost propagator $G(p^2)$ using the fitting function
$z / ((p^2)^{1+a_G} (1 + a_1 p^2 + a_2 p^4))$ and data in
the interval $p^2 \leq 2/a^2$. For each
lattice volume $V$ we report the infrared
exponent $a_G$, the number of degrees of
freedom ($d.o.f.$) and the value of $\chi^2/d.o.f.$}
\label{tab:fit_ghost}
\begin{tabular}{c c c c}   
$V$ & $\;\;\;\;a_G\;\;\;\;$ & $\chi^2 / d.o.f.$ & $ \# d.o.f.$ \\ \hline
$20^3\;$ & 0.272 (6) & 0.4 & 1 \\ 
$20^2 \times 40\;$ & 0.18 (1) & 4.9 & 6 \\ 
$20^2 \times 60\;$ & 0.154 (9) & 3.9 & 11 \\ 
$60^3\;$ & 0.21 (1) & 4.6 & 11 \\ 
$8^2 \times 64\;$ & 0.053 (7) & 1.7 & 12 \\
$8^2 \times 140\;$ & 0.025 (2)  & 1.9 & 31   \\
$12^2 \times 140\;$ & 0.060 (4) & 2.0 & 31 \\ 
$16^2 \times 140\;$ & 0.093 (6) & 2.0 & 31 \\ 
$140^3\;$ & 0.174 (7) & 8.5 & 31 \\ \hline
\end{tabular}
\end{table}

Let us also note that in Figures \ref{fig:gluon1} and \ref{fig:gluon2}
the gluon propagator at zero momentum $D(0)$ is monotonically
decreasing when considered as a function of the lattice volume $V$.
On the other hand, if one considers the symmetric lattice
$ V = 20^3 = 8000 $ and the slightly larger asymmetric lattice
$V = 8^2 \times 140 = 8960 $, one finds (see Figure \ref{fig:gluon3})
that $ D(0) $ is {\em larger} (by about $ 20 \% $) in the asymmetric case.
Thus, an extrapolation of $D(0)$ 
to infinite volume using data from asymmetric lattices
is most likely also affected by systematic effects.

We also tried a fit to the data using the fitting functions 
considered in Ref.\ \cite{Silva:2005hb} (see results in
Table \ref{tab:fit_gluon}). We see that the IR exponent
$ a_D $ usually increases with the lattice volume,
in agreement with the results reported in Table 2 of Ref.\ \cite{Silva:2005hb}.
However, it is clear from Figure \ref{fig:extrap} that the dependence
of $ a_D $ on $ 1/ V $ is not simple. Indeed, $ a_D $ looks almost
constant when considering $ V = 20^2 \times 40 $ and $ 20^2 \times 60 $,
but it gets a larger value when using the symmetric --- and much larger ---
lattice volumes $ V = 60^3 $ and $140^3$. A smaller value for $ a_D $ 
is also found if one tries an extrapolation as a function of $ 1/V $ for the
results obtained using the three asymmetric lattices $ V = N_s^2 \times 140 $.
Again, it could be difficult to have systematic
errors under control when trying an extrapolation using only asymmetric lattices.

Strong systematic effects can also be observed in the ghost case. The propagator
is less IR divergent when one considers asymmetric lattices
(see Figures \ref{fig:ghost1} and \ref{fig:ghost3}). As a consequence, the IR
exponent $ a_G $ is also smaller (see Table \ref{tab:fit_ghost})
for asymmetric lattices. This is the case either
when comparing an asymmetric lattice to a symmetric one with almost equal
lattice volume (i.e.\ $V = 8^2 \times 140$ and $V = 20^3$) or when considering
an asymmetric lattice and a symmetric one with the same largest lattice side
(e.g.\ $V = 16^2 \times 140$ and $V = 140^3$). Moreover, the dependence on
the lattice volume is very different in the two cases. Indeed, for
symmetric lattices (e.g.\ $V = 20^3$, $60^3$ and $140^3$),
the IR exponent decreases as the lattice volume $V$ increases
(at fixed $\beta$). This is in agreement with the result obtained in the
$4d$ $SU(2)$ case (see Table 4 and Figure 3 in Ref.\ \cite{Cucchieri:1997dx}).
On the other hand, $ a_G $
increases as a function of V for the lattices $ V = N_s^2 \times 140 $
with increasing $N_s$. Thus, there is no simple relation between $ a_G $ and
$ V $ and an extrapolation in $ V $ using asymmetric lattices is
probably of difficult interpretation.

Using the above results we can evaluate the quantity $a_D - 2 a_G - 1/2$, which
is zero using DSE's \cite{Zwanziger:2001kw,Lerche:2002ep}. We find 
$a_D - 2 a_G - 1/2 = 0.21(2)$ for $V = 60^3$ and $0.28(2)$ for $V = 140^3$.
For the asymmetric lattices this value is somewhat larger.
However, let us note that these results depend on the choice of the fitting
function. Moreover, the null value for $a_D - 2 a_G - 1/2$ should be obtained
only in the infinite-volume and in the continuum limit. We will analyze
these limits in a future study.

In this work we did not do a careful study of Gribov-copy effects
\cite{Cucchieri:1997dx,Cucchieri:1997ns,Giusti:2001xf,Mandula:nj,
Silva:2004bv,Nakajima:2004vc,Sternbeck:2004xr,Sternbeck:2005tk,Lokhov:2005ra,
Bogolubsky:2005wf},
since we are interested in possible systematic effects
due to the use of asymmetric lattices \footnote{Note that evidences
of Gribov-copy effects in lattice Landau gauge
have been found for gluon and ghost propagators,
the horizon tensor, the smallest eigenvalue of the
Faddeev-Popov matrix, the Kugo-Ojima parameter and $\alpha_s(p)$
(defined using gluon and ghost propagators).}.
Nevertheless, for the gluon propagator
and $V = 20^3$ and $8^2 \times 140$, we compare results
obtained using
a standard gauge-fixing algorithm to data obtained with the
so-called smearing method \cite{Hetrick:1997yy,Cucchieri:2004sq}.
(In this case we considered 2000
configurations for each lattice volume and each gauge-fixing method.)
We found that for
$ V = 20^3 $ the value of $D(0)$ is $0.873(9)$ when using
the smearing method, while one gets $0.906(9)$ with a standard gauge-fixing method.
The two results differ by several standard deviations and
the propagator is smaller when using the smearing method, i.e.\ as
one gets closer to the so-called fundamental modular region.
On the contrary, for $ V = 8^2 \times 140 $ we found
$D(0) = 1.183(15)$ with smearing and $D(0) = 1.152(15)$ without smearing.
In this case the relative difference between the two cases is smaller
and the result using smearing is {\em larger} than in the standard case.
This seems to be a systematic effect, which we observe for the first nine
smaller momenta, when considering the asymmetric lattice.
We believe that this issue deserves a more careful analysis.

\section{Conclusions}

We have compared data for gluon and ghost
propagators (in minimal Landau gauge)
obtained using symmetric and asymmetric
lattices. We find, for both propagators, clear evidences of systematic effects
at relatively small momenta ($p \lesssim 650 \mbox{MeV} 
\approx 1.5 \sqrt{\sigma}$). In particular, the gluon (respectively, ghost)
propagator is less suppressed (respec.\ enhanced) in the IR limit
when considering asymmetric lattices than for the case of symmetric lattices.
This implies that the estimates for the IR critical exponents $a_G$
and $ a_D $ are systematically smaller in the asymmetric case compared
to the symmetric one. Also, for the gluon propagator, the turnover
point $ M $ is significantly smaller
when considering asymmetric lattices than for the symmetric ones.
Let us recall that $M$ corresponds to the Gribov mass scale in a
Gribov-like propagator $D(p^2) = p^2 / (p^4 + M^4)$.
Finally, we have seen that the extrapolation to infinite volume of results 
obtained using asymmetric lattices is also most likely affected by systematic
errors.
We conclude that, even though using an asymmetric lattice does not modify the
qualitative behavior of the two propagators, one should be careful
in extracting quantitative information from such studies.

Our data have been obtained in the $3d$ $SU(2)$ case. However, the
behavior observed for the gluon propagator is very similar to
what is obtained in the $4d$ $SU(3)$ case 
\cite{Oliveira:2004gy,Silva:2005hd,Oliveira:2005hg,Silva:2005hb},
where of course a study of this type is more complicated.


\section*{ACKNOWLEDGMENTS}

The authors thank A. Maas for helpful comments and suggestions.
Research supported by FAPESP (Projects No.\ 00/ 05047-5).
Partial support from CNPq is also acknowledged.


\end{document}